\documentclass{article}
\usepackage{graphicx}
\usepackage[round]{natbib}
\usepackage{epsfig}
\usepackage{longtable}
\usepackage{subfigure}
\title{
Statistical Modelling of the Relationship Between Main
Development Region Sea Surface Temperature and \emph{Landfalling} Atlantic Basin
Hurricane Numbers
 }

\author{
Roman Binter (RMS and LSE)\\
Stephen Jewson (RMS) \footnote{\emph{Correspondence email}: \texttt{stephen.jewson@rms.com}}\\
Shree Khare (RMS)\\}

\begin{document}
\maketitle

\begin{abstract}
We are building a hurricane number prediction scheme that relies, in part, on statistical modelling of the
empirical relationship between Atlantic sea surface temperatures and landfalling hurricane numbers.
We test out a number of simple statistical models for that relationship, using data from 1900 to 2005
and data from 1950 to 2005, and for both all hurricane numbers and intense hurricane numbers.
The results are very different from the corresponding analysis for basin hurricane numbers.
\end{abstract}

\section{Introduction}


We are interested in developing practical methods for the
prediction of the distribution of the number of hurricanes that
might make landfall in the US over the next 5 years. One possible
way to make such predictions is via a 2-step method that involves
predicting main development region (MDR) sea surface temperature
(SST), and then predicting landfalling hurricane numbers as a
function of the MDR SST. The first step of predicting MDR SST has
been considered in~\citet{j92} and~\citet{e20}. This paper investigates the
second step, and considers statistical relations that one might
use to model the relationship between MDR SST and the number of
landfalls. This paper closely follows an earlier paper~\citep{e04a} in
which we modelled the relationship between MDR SST and the number
of hurricanes in the basin. The data and the models we use, and
the format of the results we present, are all taken from that
paper. Readers should refer to that paper for further details
including a short discussion, with references, giving an overview of the
physical relationship between SST and hurricanes.

The rest of this article proceeds as follows: in
section~\ref{results} we show the results from our tests on the
landfalling hurricane data, and in section~\ref{summary} we
discuss what we have found.

\section{Results}\label{results}

We now present results from our comparisons of the ability of various
statistical models to represent the relationship between MDR SST
and landfalling hurricane numbers.
First we consider models for the total number of landfalling hurricanes, for the periods 1900-2005 and
1950-2005, and then we consider models for intense landfalling hurricane numbers for the same two periods.

\subsection{All hurricanes, 1900-2005}

The first results we present are based on all landfalling
hurricanes, and data from 1900 to 2005.

The scatter plot shown in figure~1 shows the number of hurricanes
versus the SST during this period. The picture is dramatically
different from what we saw when we considered the relationship
between SST and the number of hurricanes in the basin in~\citet{e04a}. In that
case there was a clear positive relationship. In this case, there
is, at least \emph{prima facie}, no relationship at all. Table~1
shows that the linear correlation is 0.16 and the rank
correlation is 0.12.

Table~2 shows the score comparisons for the six models for this
data set, and table~4 shows the p-values for the pairwise
comparison of these models. The best model, in terms of
out-of-sample RMSE performance, is the exponential negative
binomial model, but the score is only slightly better than the trivial flat
poisson model with has no relationship between landfalling numbers
and SST. The point-wise comparisons shows that the non-trivial
models are not statistically distinguishable from the trivial
model.

As far as the log-likelihood scores in table~2 are concerned, we
find that the flat poisson model defeats the linear and damped
linear normal models in a statistically significant way. The
remaining models do not beat the flat-line model in a
statistically significant way.

In table~3 we see that the slope parameters of all the non-trivial
models are not significantly different from zero. In other words,
based on these parameter estimates and standard errors, we
certainly couldn't reject a null-hypothesis that there is no
relation at all between MDR SST and the number of landfalling
hurricanes.

The damping parameter in the damped linear trend model is  much
less than 1 (0.73), in response to the weak (or perhaps
non-existent) signal that we are trying to identify.

The linear relationships between SST and hurricane numbers, for
what they are worth, give a sensitivity of between 0.64 and 0.69
hurricanes per degree.

In summary: we don't see any indication of a relationship between MDR and SST and landfalling hurricane numbers.

\subsection{All hurricanes, 1950-2005}

Given the lack of a significant relationship between SST and
hurricane numbers on the data from 1900 to 2005, it is interesting
to see if we can find one using the more recent data. On the one
hand, the more recent data is more accurate, and so it might be
more likely we can detect a relationship. On the other hand, using
less data will make it even harder to estimate the parameters of
the models.

Table~6 shows that all the non-trivial models defeat the flat
poisson model. The results for linear normal, damped linear normal
and linear poisson are statistically significant.

As far as the log-likelihood scores are concerned, the linear
normal and damped linear normal models are defeated by the flat
poisson model in a statistically significant way. The linear
poisson model defeats the trivial model in a statistically
significant way (for RMSE). However, based on these results,
it's hard to conclude definitively that the linear poisson model
is best. For instance, in the pairwise comparisons for RMSE and
log-likelihood, the exponential poisson model also defeats the linear
poisson model in a statistically significant way.

Once again the slope parameters in the non-trivial models are all
indistinguishable from zero. The linear relationships now have
slopes between 0.99 and 1.2, but with standard errors of up to
0.63.

Overall, there is at least now \emph{some} statistical evidence of a relationship
between SST and landfalling hurricane numbers, although the parameters
of the relationship are very poorly estimated.

\subsection{Intense hurricanes}

We have only just found a relationship between MDR SSTs and the
total number of landfalling hurricanes. Could there be a
relationship between MDR SSTs and the number of \emph{intense}
landfalling hurricanes, for which there is even less data? The scatter plots (figures 7 and 8)
suggest that there is no such relationship. Our statistical
results (tables~10 to 17) confirm that none of the non-trivial
models defeat the flat trivial model in a statistically
significant way for both data sets 1900-2005 and 1950-2005. In
fact, for the 1950-2005 data set, the flat-line model defeats the
linear and damped linear normal model in a statistically
significant way in the probabilistic comparison.

\section{Summary}\label{summary}

We have investigated whether there is a statistical relationship
between MDR SST and the number of hurricanes making landfall in
the US. In previous work we've seen a strong relationship between
MDR SST and the number of Atlantic basin hurricanes, and so our
\emph{a priori} assumption is that there must be some relationship
for landfall numbers as well. Our analysis, however, finds only a weak
relationship for total hurricane numbers and no relationship at all
for intense hurricane numbers.


Why could this be? We see two possible explanations:
\begin{itemize}

    \item There \emph{is} a physical relationship between SST and landfalling hurricane numbers, but this relationship
    is mostly obscured by the signal-to-noise ratio, which is very poor because there are so few landfalling
    hurricanes.

    \item Even though there is a strong physical relationship between SST and basin hurricane numbers, there is only
    a weak physical relationship between SST and landfalling hurricane numbers. The effects of SST conspire to change
    the proportion of hurricanes that make landfall in such a way that the effects that are seen in the basin
    numbers almost disappear when we consider the landfalls.
    For instance, higher SSTs may mean higher numbers of hurricanes in the basin (on average), but they may also mean
    a lower proportion making landfall,
    and these two effects may combine in such a way that the actual number making landfall remains the same.

\end{itemize}


And of course reality may be a combination of these two effects.

We are investigating both of these possibilities. W.r.t. the
first, we are considering simple statistical systems to see
whether this is really the behaviour that we would expect under
idealized assumptions. W.r.t. the second, we are performing
statistical tests to investigate the hypothesis that the
proportion of hurricanes making landfall really does vary in such
a way as to mask the effects of SST on numbers of hurricanes.

\bibliography{arxiv}

\begin{thebibliography}{3}
\providecommand{\natexlab}[1]{#1}
\providecommand{\url}[1]{\texttt{#1}}
\expandafter\ifx\csname urlstyle\endcsname\relax
  \providecommand{\doi}[1]{doi: #1}\else
  \providecommand{\doi}{doi: \begingroup \urlstyle{rm}\Url}\fi

\bibitem[Binter et~al.(2006)Binter, Jewson, and Khare]{e04a}
R~Binter, S~Jewson, and S~Khare.
\newblock {Statistical modelling of the relationship between Main Development
  Region Sea Surface Temperature and Atlantic Basin hurricane numbers}.
\newblock \emph{arXiv:physics/0701170}, 2006.
\newblock RMS Internal Report E04a.

\bibitem[Laepple et~al.(2006)Laepple, Jewson, Meagher, O'Shay, and Penzer]{e20}
T~Laepple, S~Jewson, J~Meagher, A~O'Shay, and J~Penzer.
\newblock {Five-year ahead prediction of Sea Surface Temperature in the
  Tropical Atlantic: a comparison of simple statistical methods}.
\newblock \emph{arXiv:physics/0701162}, 2006.

\bibitem[Meagher and Jewson(2006)]{j92}
J~Meagher and S~Jewson.
\newblock {Year ahead prediction of hurricane season SST in the tropical
  Atlantic}.
\newblock \emph{arXiv:physics/0606185}, 2006.

\end{thebibliography}

\newpage

\begin{table}[h!]
\begin{center}
\caption{Linear and Rank Correlations} {\small
\begin{tabular}{|c|c|c|}
\hline
 & Linear Correlation & Rank Correlation \\
\hline
1900 - 2005 Landfall vs SST & 0.16 & 0.12 \\
1950 - 2005 Landfall vs SST & 0.25 & 0.16 \\
1900 - 1949 Landfall vs SST & 0.17 & 0.25 \\
1900 - 2005 Intense Landfall vs SST & 0.28 & 0.24 \\
1950 - 2005 Intense Landfall vs SST & 0.25 & 0.09 \\
\hline
\end{tabular}
}
\end{center}
\end{table}

\newpage

\begin{table}[h!]
\begin{center}
\caption{RMSE comparison 1900 - 2005 Landfall vs SST} {\small
\begin{tabular}{|c|c|c|c|c|c|c|}
\hline
 & model name & RMSE (in) & RMSE (out) & 100-100*RMSE/RMSEconst & LL (in) & LL (out) \\
\hline
model 1 & Flat Poisson & 1.399 & 1.412 & 0 & -1.655 & -1.666 \\
model 2 & Linear Normal & 1.381 & 1.411 & 0.165 & -1.841 & -1.864 \\
model 3 & Damped Linear Normal & 1.382 & 1.415 & -0.359 & -1.842 & -1.866 \\
model 4 & Linear Poisson & 1.381 & 1.41 & 0.372 & -1.642 & -1.664 \\
model 5 & Exponential Poisson & 1.379 & 1.411 & 0.264 & -1.641 & -1.664 \\
model 6 & Exponential Neg. Bin. & 1.379 & 1.387 & 3.577 & -1.638 & -1.649 \\
\hline
\end{tabular}
}
\end{center}
\end{table}
\begin{table}[h!]
\begin{center}
\caption{Model parameters incl. out of sample RMSE 1900 - 2005
Landfall vs SST} {\small
\begin{tabular}{|c|c|c|c|c|c|c|c|c|}
\hline
 & $\hat{\alpha}$ & s.e. & $\hat{\beta}$ & s.e. & $k$ & cov & corr & RMSE (out of sample) \\
\hline
model 1 & 0.541 & 0.074 &  &  &  &  &  & 1.412 \\
model 2 & 1.717 & 0.135 & 0.686 & 0.415 &  & 0 & 0 & 1.411 \\
model 3 & 1.717 &  & 0.502 &  & 0.732 &  &  & 1.415 \\
model 4 & 1.717 & 0.127 & 0.645 & 0.39 &  & 0.006 & 0.123 & 1.41 \\
model 5 & 0.532 & 0.075 & 0.395 & 0.225 &  & -0.002 & -0.128 & 1.411 \\
model 6 & 0.532 & 0.078 & 0.391 & 0.236 &  & -0.002 & -0.115 & 1.387 \\
\hline
\end{tabular}
}
\end{center}
\end{table}
\begin{table}[h!]
\begin{center}
\caption{Winning count for particular model 1900 - 2005 Landfall
vs SST} {\small
\begin{tabular}{|c|c|c|c|c|c|c|}
\hline
 & model 1 & model 2 & model 3 & model 4 & model 5 & model 6 \\
\hline
model 1 & 0 (1) & 44 (0.897) & 44 (0.897) & 44 (0.897) & 44 (0.897) & 43 (0.928) \\
model 2 & 56 (0.143) & 0 (1) & 53 (0.314) & 52 (0.385) & 57 (0.103) & 45 (0.857) \\
model 3 & 56 (0.143) & 47 (0.752) & 0 (1) & 47 (0.752) & 46 (0.809) & 49 (0.615) \\
model 4 & 56 (0.143) & 48 (0.686) & 53 (0.314) & 0 (1) & 50 (0.539) & 45 (0.857) \\
model 5 & 56 (0.143) & 43 (0.928) & 54 (0.248) & 50 (0.539) & 0 (1) & 49 (0.615) \\
model 6 & 57 (0.103) & 55 (0.191) & 51 (0.461) & 55 (0.191) & 51 (0.461) & 0 (1) \\
\hline
\end{tabular}
}
\end{center}
\end{table}
\begin{table}[h!]
\begin{center}
\caption{Winning count (LL) for particular model 1900 - 2005
Landfall vs SST} {\small
\begin{tabular}{|c|c|c|c|c|c|c|}
\hline
 & model 1 & model 2 & model 3 & model 4 & model 5 & model 6 \\
\hline
model 1 & 0 (1) & 76 (0) & 75 (0) & 44 (0.897) & 44 (0.897) & 43 (0.928) \\
model 2 & 24 (1) & 0 (1) & 53 (0.314) & 25 (1) & 25 (1) & 22 (1) \\
model 3 & 25 (1) & 47 (0.752) & 0 (1) & 22 (1) & 23 (1) & 22 (1) \\
model 4 & 56 (0.143) & 75 (0) & 78 (0) & 0 (1) & 50 (0.539) & 47 (0.752) \\
model 5 & 56 (0.143) & 75 (0) & 77 (0) & 50 (0.539) & 0 (1) & 49 (0.615) \\
model 6 & 57 (0.103) & 78 (0) & 78 (0) & 53 (0.314) & 51 (0.461) & 0 (1) \\
\hline
\end{tabular}
}
\end{center}
\end{table}
\begin{figure}[h!]
\centering {
\includegraphics[width=8cm, angle=-90]{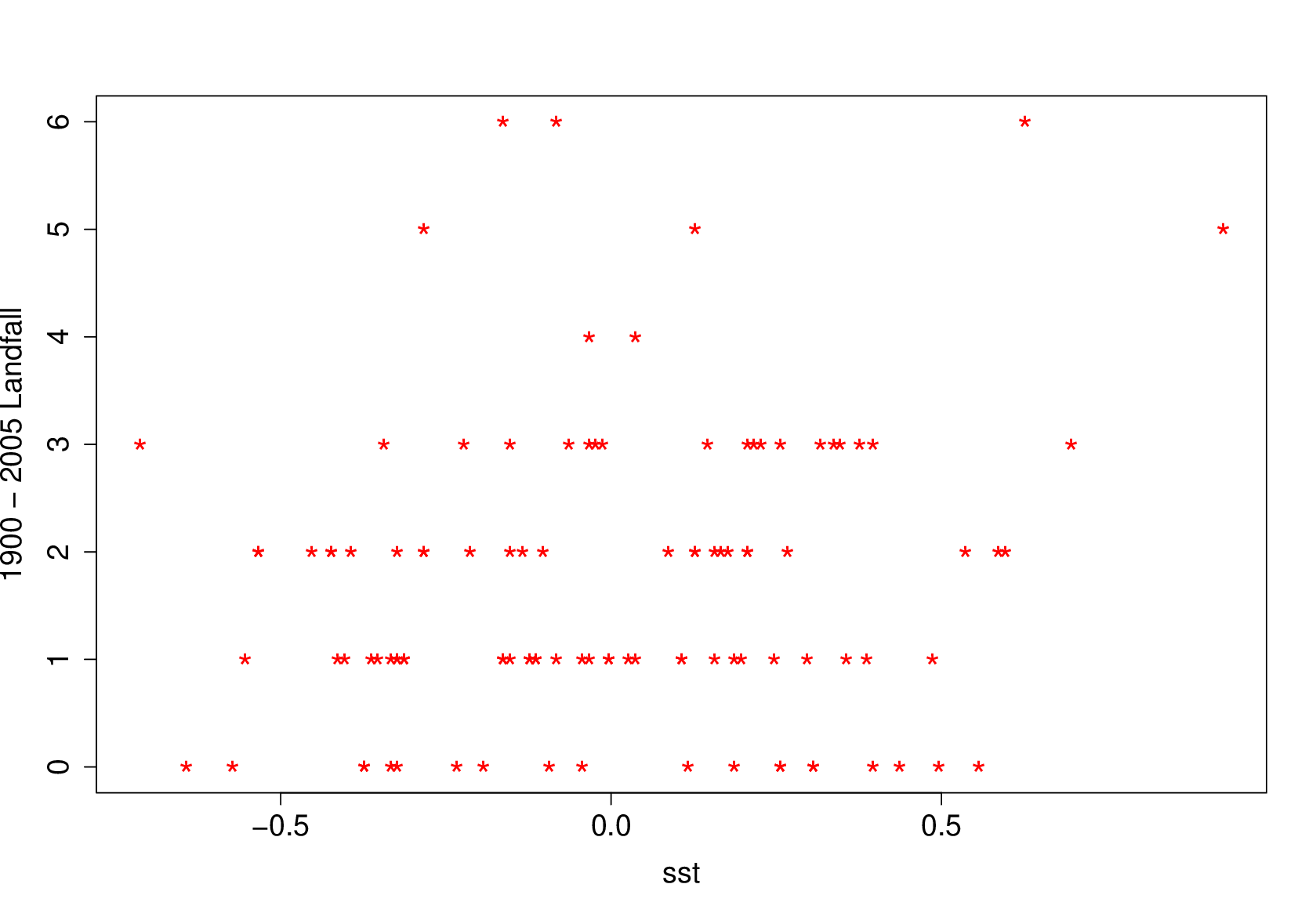}
} \caption{1900 - 2005 Landfall vs. SST}
\end{figure}

\begin{figure}[h!]
\centering {
\includegraphics[width=8cm, angle=-90]{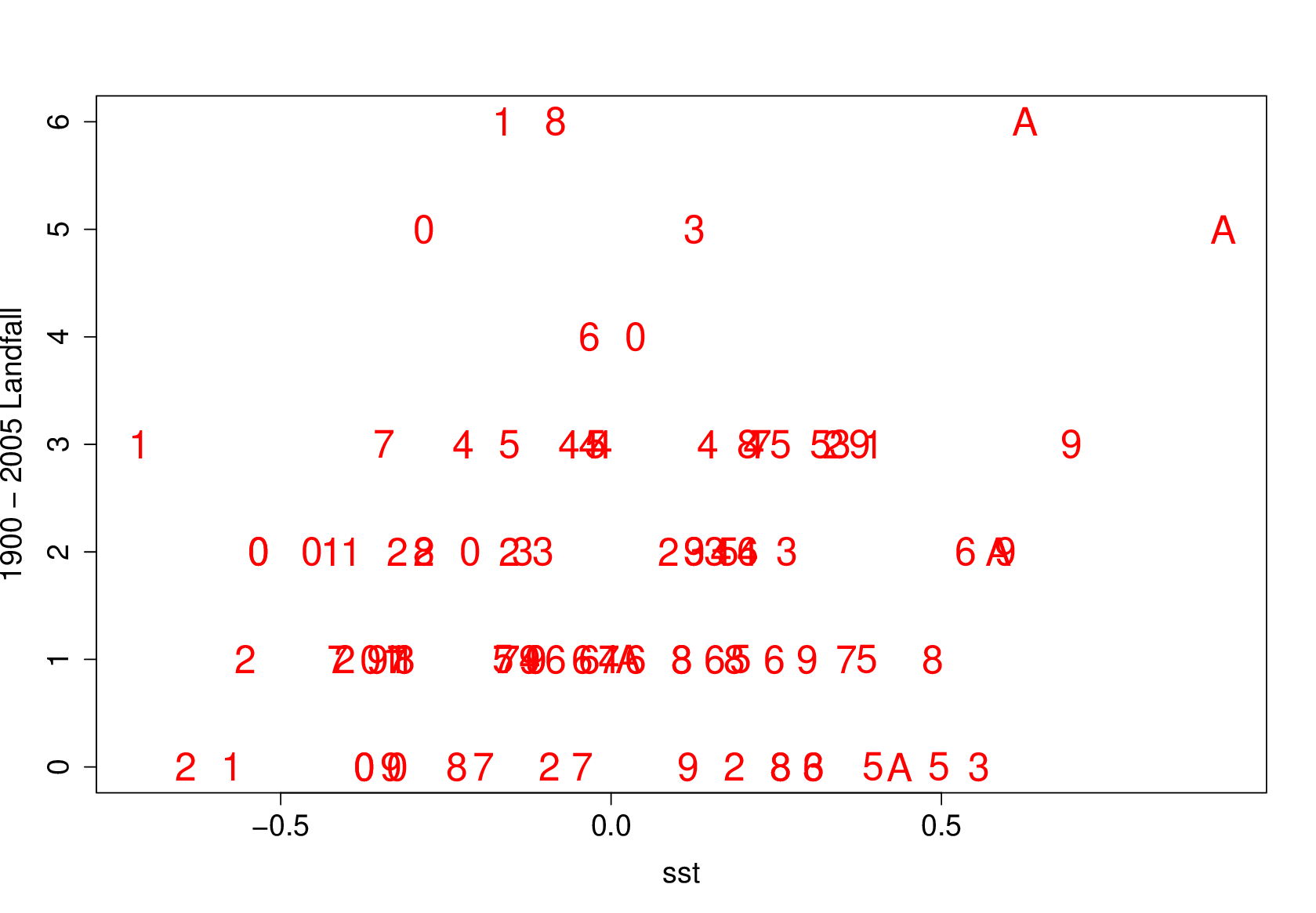}
} \caption{1900 - 2005 Landfall vs. SSTs}
\end{figure}
\clearpage

\begin{figure}[h!]
\centering {
\includegraphics[width=10cm, angle=-90]{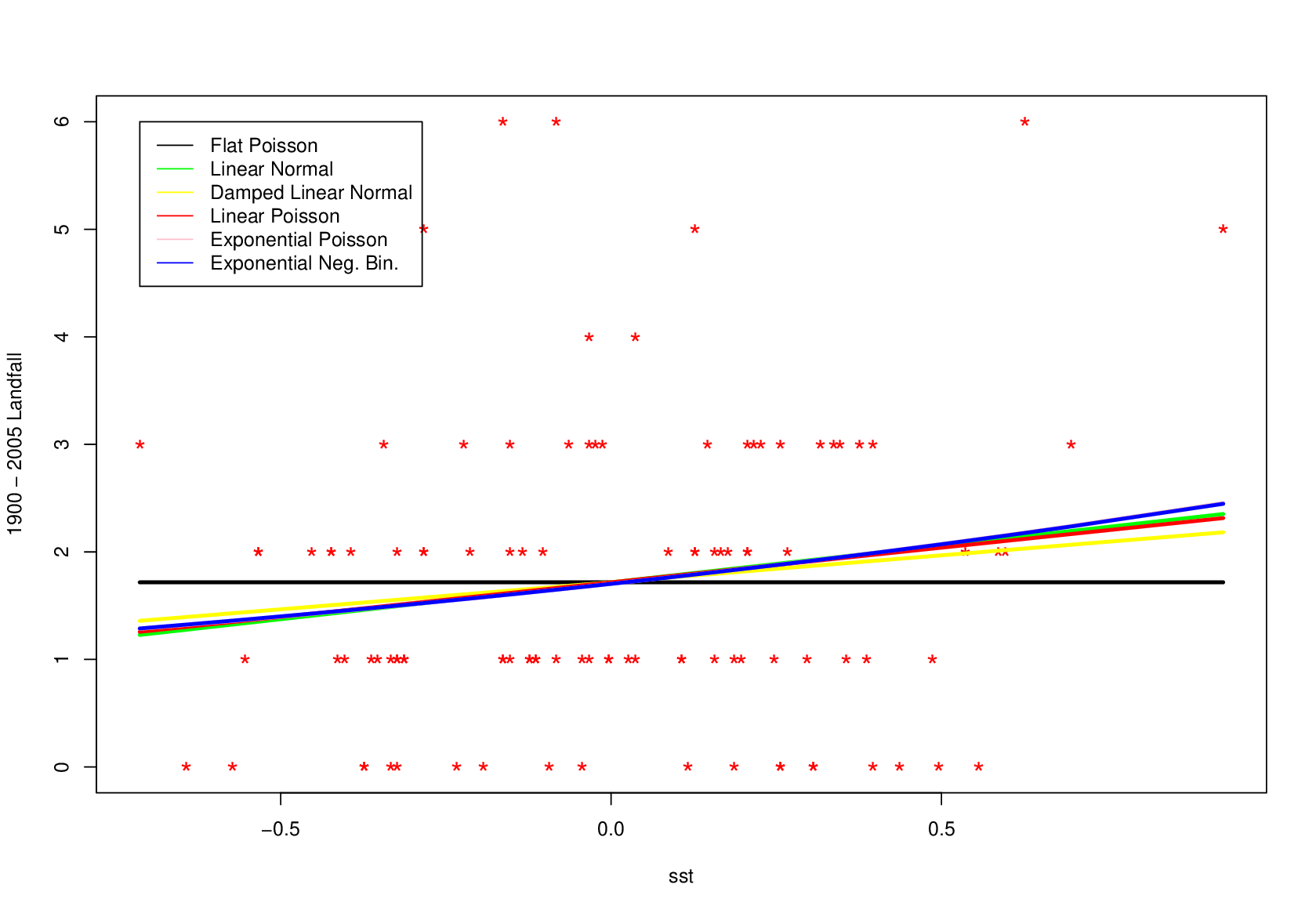}
} \caption{Fitted Lines for all Models 1900 - 2005 Landfall vs
SST}
\end{figure}
\newpage

\begin{table}[h!]
\begin{center}
\caption{RMSE comparison 1950 - 2005 Landfall vs SST constant}
{\small
\begin{tabular}{|c|c|c|c|c|c|c|}
\hline
 & model name & RMSE (in) & RMSE (out) & 100-100*RMSE/RMSEconst & LL (in) & LL (out) \\
\hline
model 1 & Flat Poisson & 1.424 & 1.45 & 0 & -1.637 & -1.66 \\
model 2 & Linear Normal & 1.379 & 1.441 & 1.24 & -1.852 & -1.91 \\
model 3 & Damped Linear Normal & 1.381 & 1.449 & 0.057 & -1.854 & -1.914 \\
model 4 & Linear Poisson & 1.38 & 1.435 & 2.023 & -1.604 & -1.65 \\
model 5 & Exponential Poisson & 1.366 & 1.431 & 2.528 & -1.598 & -1.649 \\
model 6 & Exponential Neg. Bin. & 1.367 & 1.361 & 11.935 & -1.592 & -1.615 \\
\hline
\end{tabular}
}
\end{center}
\end{table}
\begin{table}[h!]
\begin{center}
\caption{Model parameters incl. out of sample RMSE 1950 - 2005
Landfall vs SST constant} {\small
\begin{tabular}{|c|c|c|c|c|c|c|c|c|}
\hline
 & $\hat{\alpha}$ & s.e. & $\hat{\beta}$ & s.e. & $k$ & cov & corr & RMSE (out of sample) \\
\hline
model 1 & 0.463 & 0.106 &  &  &  &  &  & 1.45 \\
model 2 & 1.589 & 0.188 & 1.18 & 0.622 &  & 0 & 0 & 1.441 \\
model 3 & 1.589 &  & 0.923 &  & 0.783 &  &  & 1.449 \\
model 4 & 1.589 & 0.168 & 0.991 & 0.562 &  & 0.018 & 0.187 & 1.435 \\
model 5 & 0.439 & 0.108 & 0.72 & 0.342 &  & -0.008 & -0.213 & 1.431 \\
model 6 & 0.44 & 0.116 & 0.696 & 0.37 &  & -0.008 & -0.179 & 1.361 \\
\hline
\end{tabular}
}
\end{center}
\end{table}
\begin{table}[h!]
\begin{center}
\caption{Winning count for particular model 1950 - 2005 Landfall
vs SST} {\small
\begin{tabular}{|c|c|c|c|c|c|c|}
\hline
 & model 1 & model 2 & model 3 & model 4 & model 5 & model 6 \\
\hline
model 1 & 0 (1) & 38 (0.978) & 38 (0.978) & 38 (0.978) & 39 (0.959) & 41 (0.93) \\
model 2 & 62 (0.041) & 0 (1) & 57 (0.175) & 57 (0.175) & 46 (0.748) & 45 (0.825) \\
model 3 & 62 (0.041) & 43 (0.886) & 0 (1) & 39 (0.959) & 43 (0.886) & 50 (0.553) \\
model 4 & 62 (0.041) & 43 (0.886) & 61 (0.07) & 0 (1) & 38 (0.978) & 50 (0.553) \\
model 5 & 61 (0.07) & 54 (0.344) & 57 (0.175) & 62 (0.041) & 0 (1) & 48 (0.656) \\
model 6 & 59 (0.114) & 55 (0.252) & 50 (0.553) & 50 (0.553) & 52 (0.447) & 0 (1) \\
\hline
\end{tabular}
}
\end{center}
\end{table}
\begin{table}[h!]
\begin{center}
\caption{Winning count (LL) for particular model 1950 - 2005
Landfall vs SST} {\small
\begin{tabular}{|c|c|c|c|c|c|c|}
\hline
 & model 1 & model 2 & model 3 & model 4 & model 5 & model 6 \\
\hline
model 1 & 0 (1) & 79 (0) & 79 (0) & 38 (0.978) & 39 (0.959) & 41 (0.93) \\
model 2 & 21 (1) & 0 (1) & 57 (0.175) & 23 (1) & 23 (1) & 21 (1) \\
model 3 & 21 (1) & 43 (0.886) & 0 (1) & 23 (1) & 23 (1) & 23 (1) \\
model 4 & 62 (0.041) & 77 (0) & 77 (0) & 0 (1) & 38 (0.978) & 48 (0.656) \\
model 5 & 61 (0.07) & 77 (0) & 77 (0) & 62 (0.041) & 0 (1) & 50 (0.553) \\
model 6 & 59 (0.114) & 79 (0) & 77 (0) & 52 (0.447) & 50 (0.553) & 0 (1) \\
\hline
\end{tabular}
}
\end{center}
\end{table}

\begin{figure}[h!]
\centering {
\includegraphics[width=8cm, angle=-90]{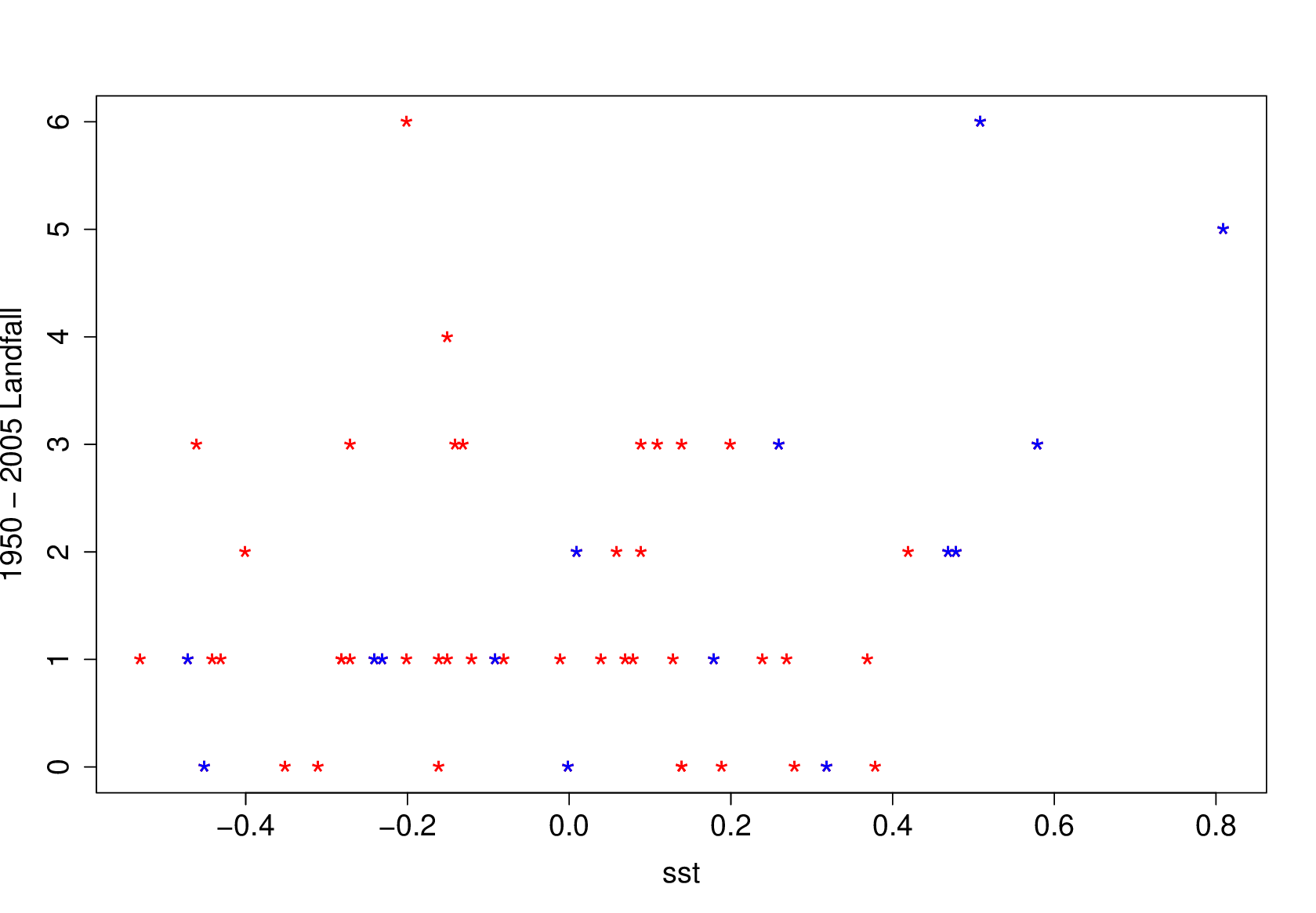}
} \caption{1950 - 2005 Landfall vs. SST}
\end{figure}

\begin{figure}[h!]
\centering {
\includegraphics[width=8cm, angle=-90]{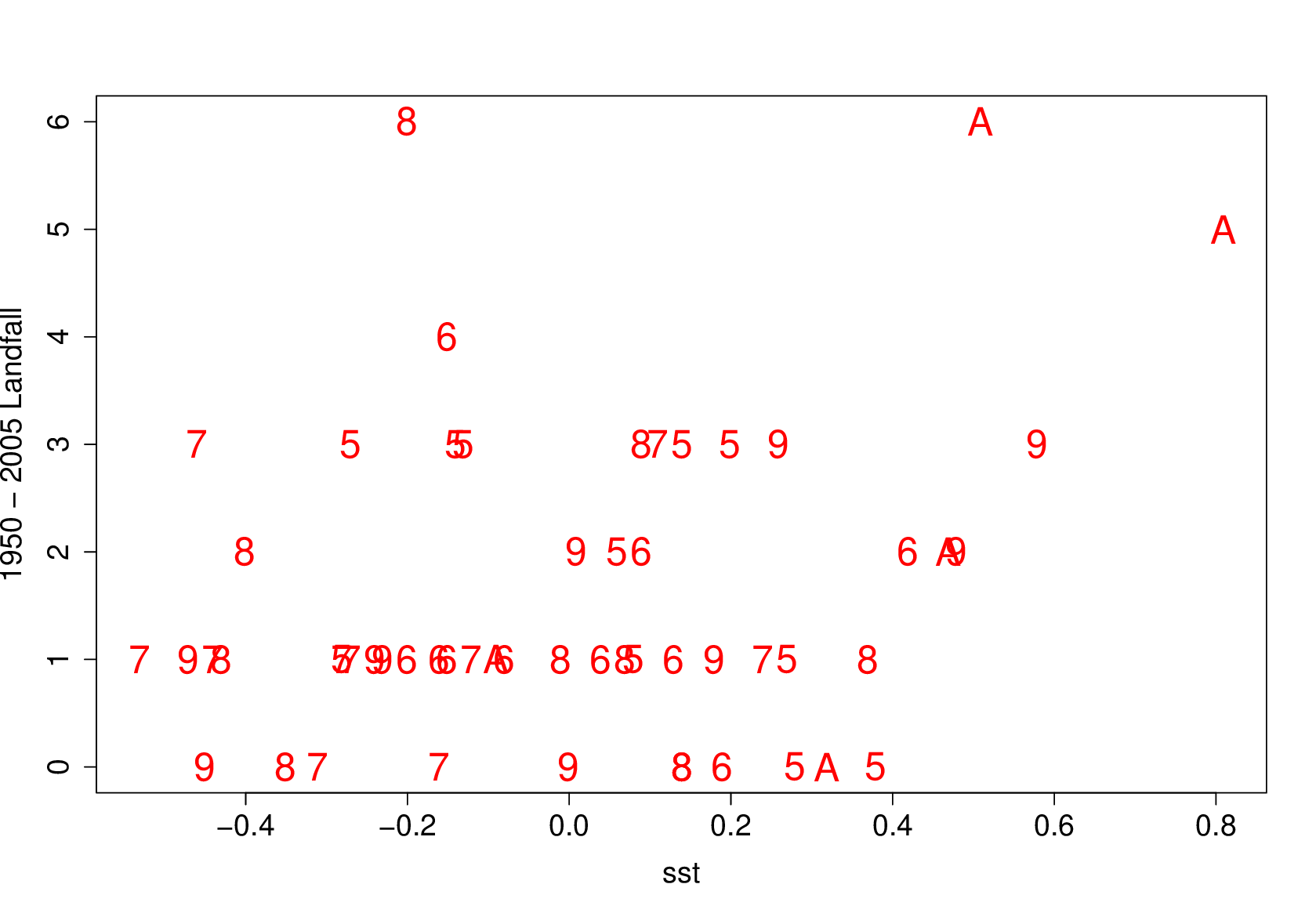}
} \caption{1950 - 2005 Landfall vs. SST}
\end{figure}
\clearpage

\begin{figure}[h!]
\centering {
\includegraphics[width=10cm, angle=-90]{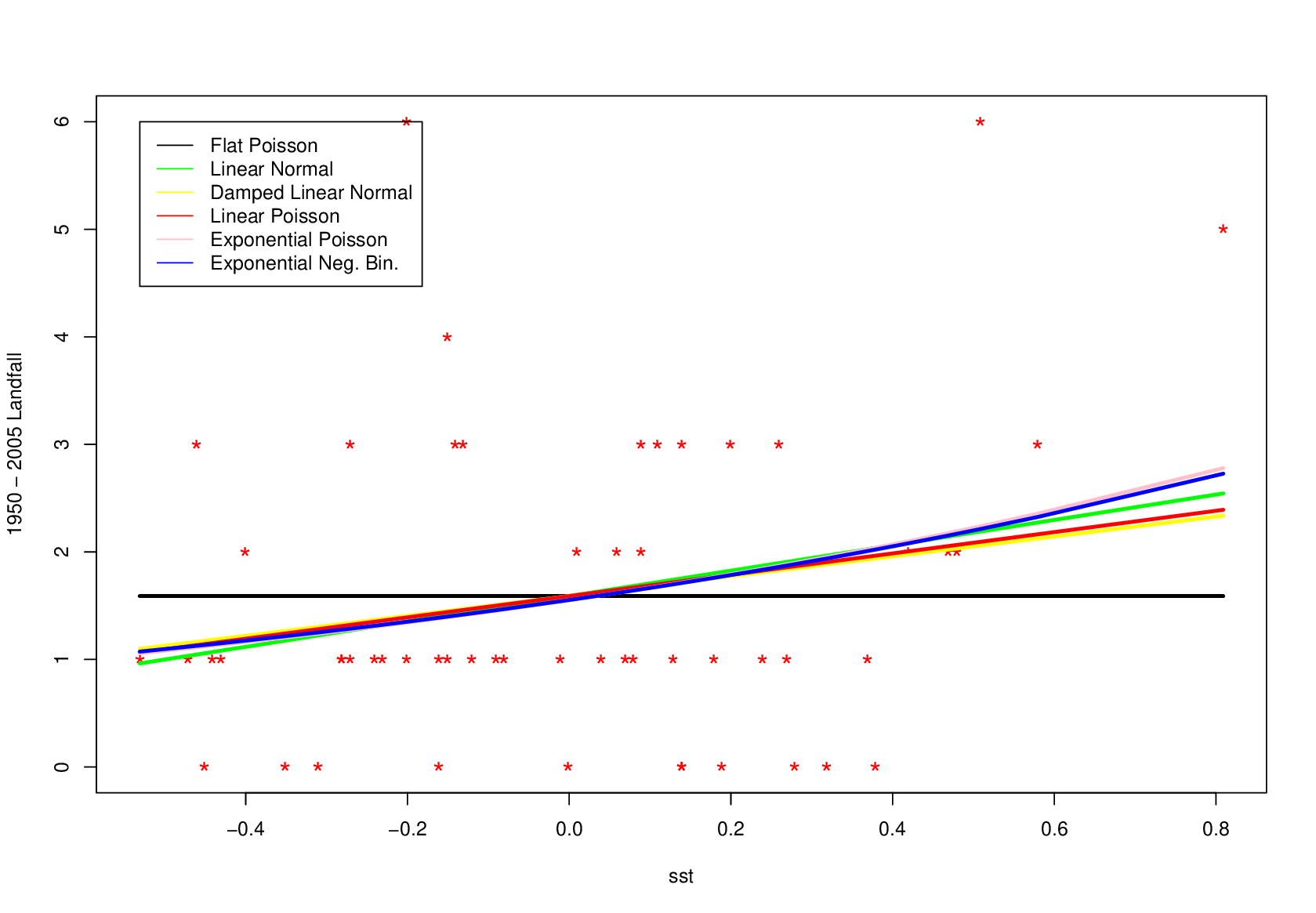}
} \caption{Fitted Lines for all Models 1950 - 2005 Landfall vs
SST}
\end{figure}

\newpage

\begin{table}[h!]
\begin{center}
\caption{RMSE comparison 1900 - 2005 Intense Landfall vs SST}
{\small
\begin{tabular}{|c|c|c|c|c|c|c|}
\hline
 & model name & RMSE (in) & RMSE (out) & 100-100*RMSE/RMSEconst & LL (in) & LL (out) \\
\hline
model 1 & Flat Poisson & 0.836 & 0.844 & 0 & -1.074 & -1.084 \\
model 2 & Linear Normal & 0.801 & 0.823 & 4.932 & -1.24 & -1.351 \\
model 3 & Damped Linear Normal & 0.802 & 0.825 & 4.538 & -1.24 & -1.356 \\
model 4 & Linear Poisson & 0.803 & 0.816 & 6.624 & -1.019 & -1.03 \\
model 5 & Exponential Poisson & 0.799 & 0.827 & 3.997 & -1.031 & -1.054 \\
model 6 & Exponential Neg. Bin. & 0.799 & 0.807 & 8.605 & -1.031 & -1.04 \\
\hline
\end{tabular}
}
\end{center}
\end{table}
\begin{table}[h!]
\begin{center}
\caption{Model parameters incl. out of sample RMSE 1950 - 2005
Intense Landfall vs SST} {\small
\begin{tabular}{|c|c|c|c|c|c|c|c|c|}
\hline
 & $\hat{\alpha}$ & s.e. & $\hat{\beta}$ & s.e. & $k$ & cov & corr & RMSE (out of sample) \\
\hline
model 1 & -0.429 & 0.12 &  &  &  &  &  & 0.844 \\
model 2 & 0.651 & 0.079 & 0.73 & 0.241 &  & 0 & 0 & 0.823 \\
model 3 & 0.651 &  & 0.659 &  & 0.902 &  &  & 0.825 \\
model 4 & 0.651 & 0.078 & 0.912 & 0.11 &  & 0.009 & 1 & 0.816 \\
model 5 & -0.494 & 0.128 & 1.094 & 0.361 &  & -0.016 & -0.338 & 0.827 \\
model 6 & -0.494 & 0.128 & 1.094 & 0.361 &  & -0.015 & -0.338 & 0.807 \\
\hline
\end{tabular}
}
\end{center}
\end{table}
\begin{table}[h!]
\begin{center}
\caption{Winning count for particular model 1900 - 2005 Intense
Landfall vs SST} {\small
\begin{tabular}{|c|c|c|c|c|c|c|}
\hline
 & model 1 & model 2 & model 3 & model 4 & model 5 & model 6 \\
\hline
model 1 & 0 (1) & 43 (0.928) & 43 (0.928) & 43 (0.928) & 44 (0.897) & 44 (0.897) \\
model 2 & 57 (0.103) & 0 (1) & 55 (0.191) & 45 (0.857) & 67 (0) & 57 (0.103) \\
model 3 & 57 (0.103) & 45 (0.857) & 0 (1) & 45 (0.857) & 59 (0.032) & 58 (0.072) \\
model 4 & 57 (0.103) & 55 (0.191) & 55 (0.191) & 0 (1) & 63 (0.004) & 59 (0.032) \\
model 5 & 56 (0.143) & 33 (1) & 41 (0.98) & 37 (0.998) & 0 (1) & 53 (0.314) \\
model 6 & 56 (0.143) & 43 (0.928) & 42 (0.951) & 41 (0.98) & 47 (0.752) & 0 (1) \\
\hline
\end{tabular}
}
\end{center}
\end{table}
\begin{table}[h!]
\begin{center}
\caption{Winning count (LL) for particular model 1900 - 2005
Intense Landfall vs SST} {\small
\begin{tabular}{|c|c|c|c|c|c|c|}
\hline
 & model 1 & model 2 & model 3 & model 4 & model 5 & model 6 \\
\hline
model 1 & 0 (1) & 50 (0.539) & 54 (0.248) & 43 (0.928) & 44 (0.897) & 44 (0.897) \\
model 2 & 50 (0.539) & 0 (1) & 55 (0.191) & 37 (0.998) & 37 (0.998) & 37 (0.998) \\
model 3 & 46 (0.809) & 45 (0.857) & 0 (1) & 37 (0.998) & 37 (0.998) & 37 (0.998) \\
model 4 & 57 (0.103) & 63 (0.004) & 63 (0.004) & 0 (1) & 63 (0.004) & 59 (0.032) \\
model 5 & 56 (0.143) & 63 (0.004) & 63 (0.004) & 37 (0.998) & 0 (1) & 54 (0.248) \\
model 6 & 56 (0.143) & 63 (0.004) & 63 (0.004) & 41 (0.98) & 46 (0.809) & 0 (1) \\
\hline
\end{tabular}
}
\end{center}
\end{table}

\begin{figure}[h!]
\centering {
\includegraphics[width=8cm, angle=-90]{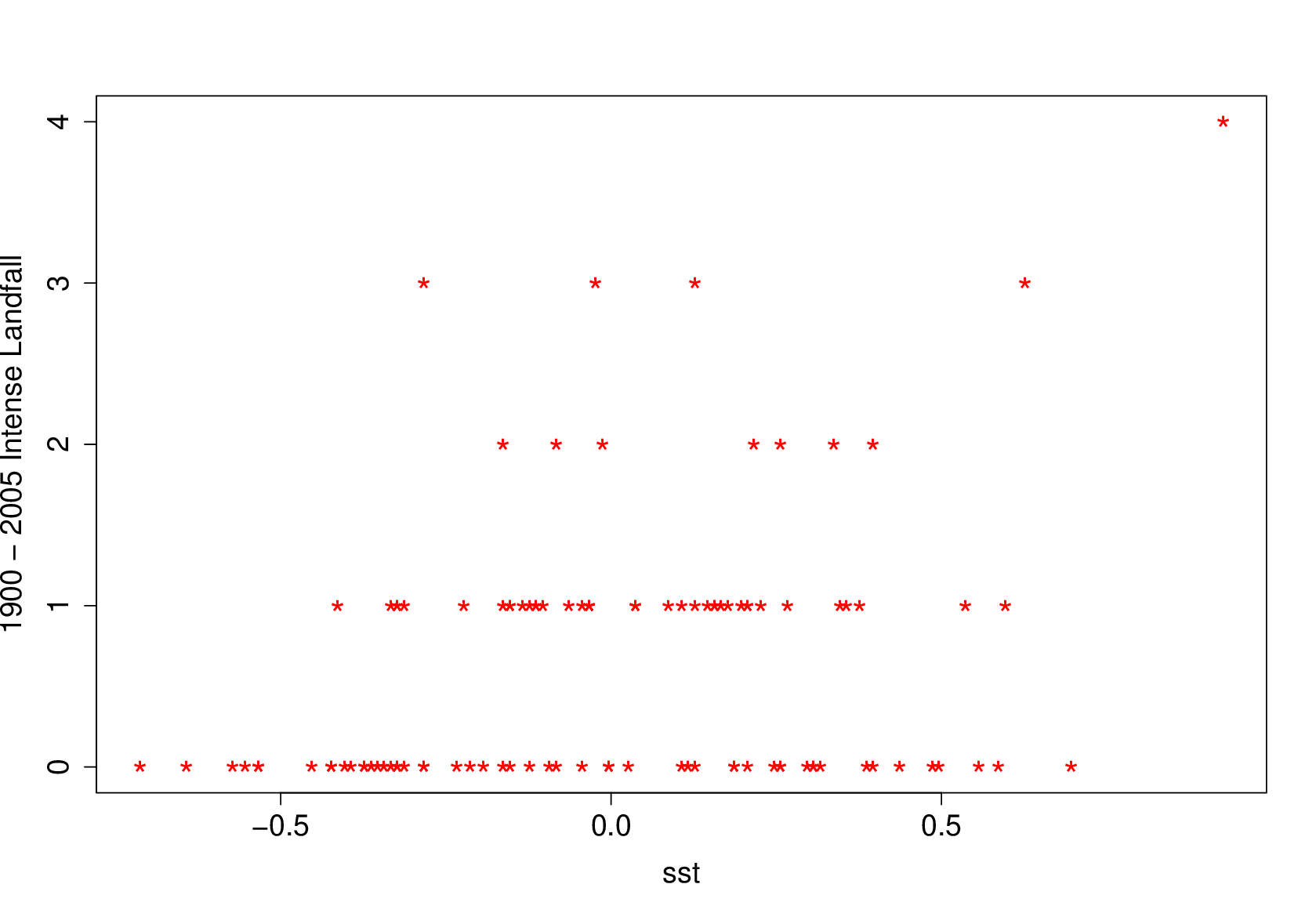}
} \caption{1900 - 2005 Intense Landfall vs SST}
\end{figure}

\begin{figure}[h!]
\centering {
\includegraphics[width=8cm, angle=-90]{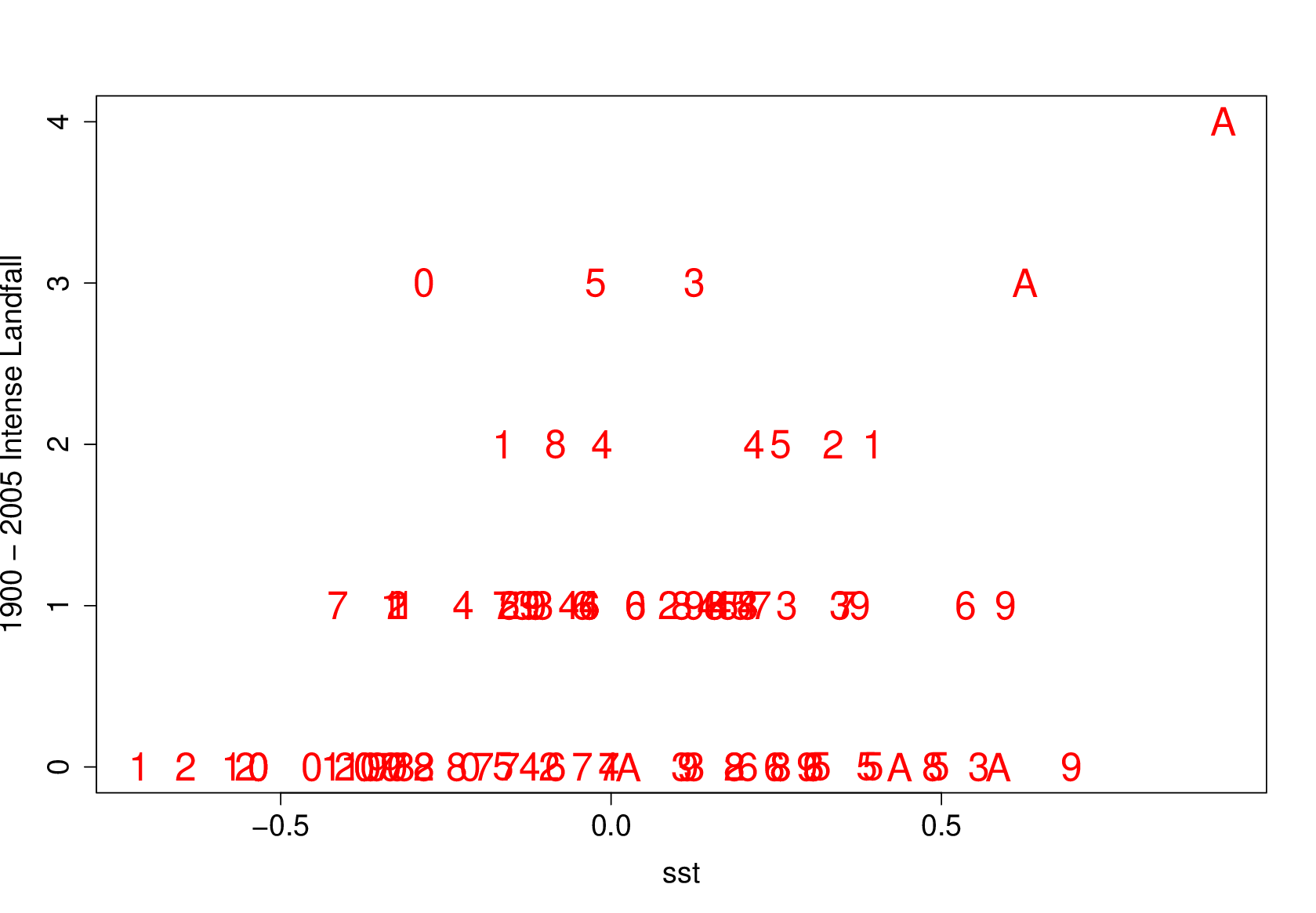}
} \caption{1900 - 2005 Intense Landfall vs SST}
\end{figure}
\clearpage

\begin{figure}[h!]
\centering {
\includegraphics[width=10cm, angle=-90]{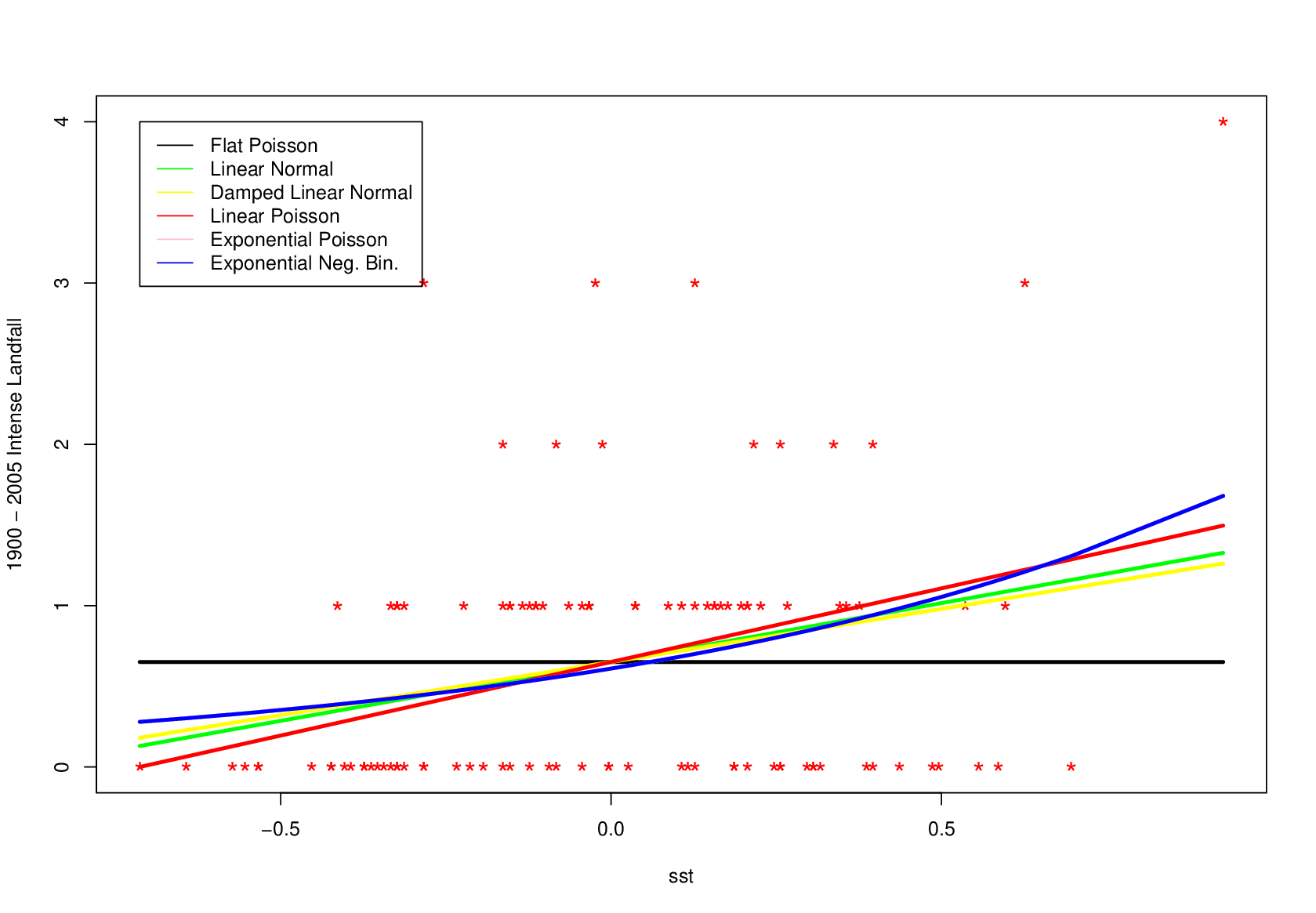}
} \caption{Fitted Lines for all Models 1900 - 2005 Intense
Landfall vs SST}
\end{figure}

\newpage

\begin{table}[h!]
\begin{center}
\caption{RMSE comparison 1950 - 2005 Intense Landfall vs SST}
{\small
\begin{tabular}{|c|c|c|c|c|c|c|}
\hline
 & model name & RMSE (in) & RMSE (out) & 100-100*RMSE/RMSEconst & LL (in) & LL (out) \\
\hline
model 1 & Flat Poisson & 0.875 & 0.891 & 0 & -1.085 & -1.107 \\
model 2 & Linear Normal & 0.848 & 0.903 & -2.692 & -1.267 & -1.514 \\
model 3 & Damped Linear Normal & 0.85 & 0.906 & -3.454 & -1.268 & -1.528 \\
model 4 & Linear Poisson & 0.848 & 0.896 & -1.211 & -1.055 & -1.112 \\
model 5 & Exponential Poisson & 0.837 & 0.903 & -2.873 & -1.05 & -1.117 \\
model 6 & Exponential Neg. Bin. & 0.837 & 0.825 & 14.307 & -1.049 & -1.067 \\
\hline
\end{tabular}
}
\end{center}
\end{table}
\begin{table}[h!]
\begin{center}
\caption{Model parameters incl. out of sample RMSE 1950 - 2005
Intense Landfall vs SST} {\small
\begin{tabular}{|c|c|c|c|c|c|c|c|c|}
\hline
 & $\hat{\alpha}$ & s.e. & $\hat{\beta}$ & s.e. & $k$ & cov & corr & RMSE (out of sample) \\
\hline
model 1 & -0.442 & 0.167 &  &  &  &  &  & 0.891 \\
model 2 & 0.643 & 0.115 & 0.71 & 0.383 &  & 0 & 0 & 0.903 \\
model 3 & 0.643 &  & 0.551 &  & 0.775 &  &  & 0.906 \\
model 4 & 0.643 & 0.107 & 0.635 & 0.35 &  & 0.011 & 0.302 & 0.896 \\
model 5 & -0.495 & 0.175 & 1.06 & 0.532 &  & -0.029 & -0.306 & 0.903 \\
model 6 & -0.493 & 0.181 & 1.03 & 0.557 &  & -0.027 & -0.278 & 0.825 \\
\hline
\end{tabular}
}
\end{center}
\end{table}
\begin{table}[h!]
\begin{center}
\caption{Winning count for particular model 1950 - 2005 Intense
Landfall vs SST} {\small
\begin{tabular}{|c|c|c|c|c|c|c|}
\hline
 & model 1 & model 2 & model 3 & model 4 & model 5 & model 6 \\
\hline
model 1 & 0 (1) & 50 (0.553) & 50 (0.553) & 50 (0.553) & 54 (0.344) & 54 (0.344) \\
model 2 & 50 (0.553) & 0 (1) & 50 (0.553) & 50 (0.553) & 54 (0.344) & 48 (0.656) \\
model 3 & 50 (0.553) & 50 (0.553) & 0 (1) & 50 (0.553) & 46 (0.748) & 52 (0.447) \\
model 4 & 50 (0.553) & 50 (0.553) & 50 (0.553) & 0 (1) & 59 (0.114) & 46 (0.748) \\
model 5 & 46 (0.748) & 46 (0.748) & 54 (0.344) & 41 (0.93) & 0 (1) & 50 (0.553) \\
model 6 & 46 (0.748) & 52 (0.447) & 48 (0.656) & 54 (0.344) & 50 (0.553) & 0 (1) \\
\hline
\end{tabular}
}
\end{center}
\end{table}
\begin{table}[h!]
\begin{center}
\caption{Winning count (LL) for particular model 1950 - 2005
Intense Landfall vs SST} {\small
\begin{tabular}{|c|c|c|c|c|c|c|}
\hline
 & model 1 & model 2 & model 3 & model 4 & model 5 & model 6 \\
\hline
model 1 & 0 (1) & 64 (0.022) & 62 (0.041) & 50 (0.553) & 54 (0.344) & 54 (0.344) \\
model 2 & 36 (0.989) & 0 (1) & 50 (0.553) & 36 (0.989) & 36 (0.989) & 39 (0.959) \\
model 3 & 38 (0.978) & 50 (0.553) & 0 (1) & 36 (0.989) & 36 (0.989) & 41 (0.93) \\
model 4 & 50 (0.553) & 64 (0.022) & 64 (0.022) & 0 (1) & 59 (0.114) & 46 (0.748) \\
model 5 & 46 (0.748) & 64 (0.022) & 64 (0.022) & 41 (0.93) & 0 (1) & 50 (0.553) \\
model 6 & 46 (0.748) & 61 (0.07) & 59 (0.114) & 54 (0.344) & 50 (0.553) & 0 (1) \\
\hline
\end{tabular}
}
\end{center}
\end{table}

\begin{figure}[h!]
\centering {
\includegraphics[width=8cm, angle=-90]{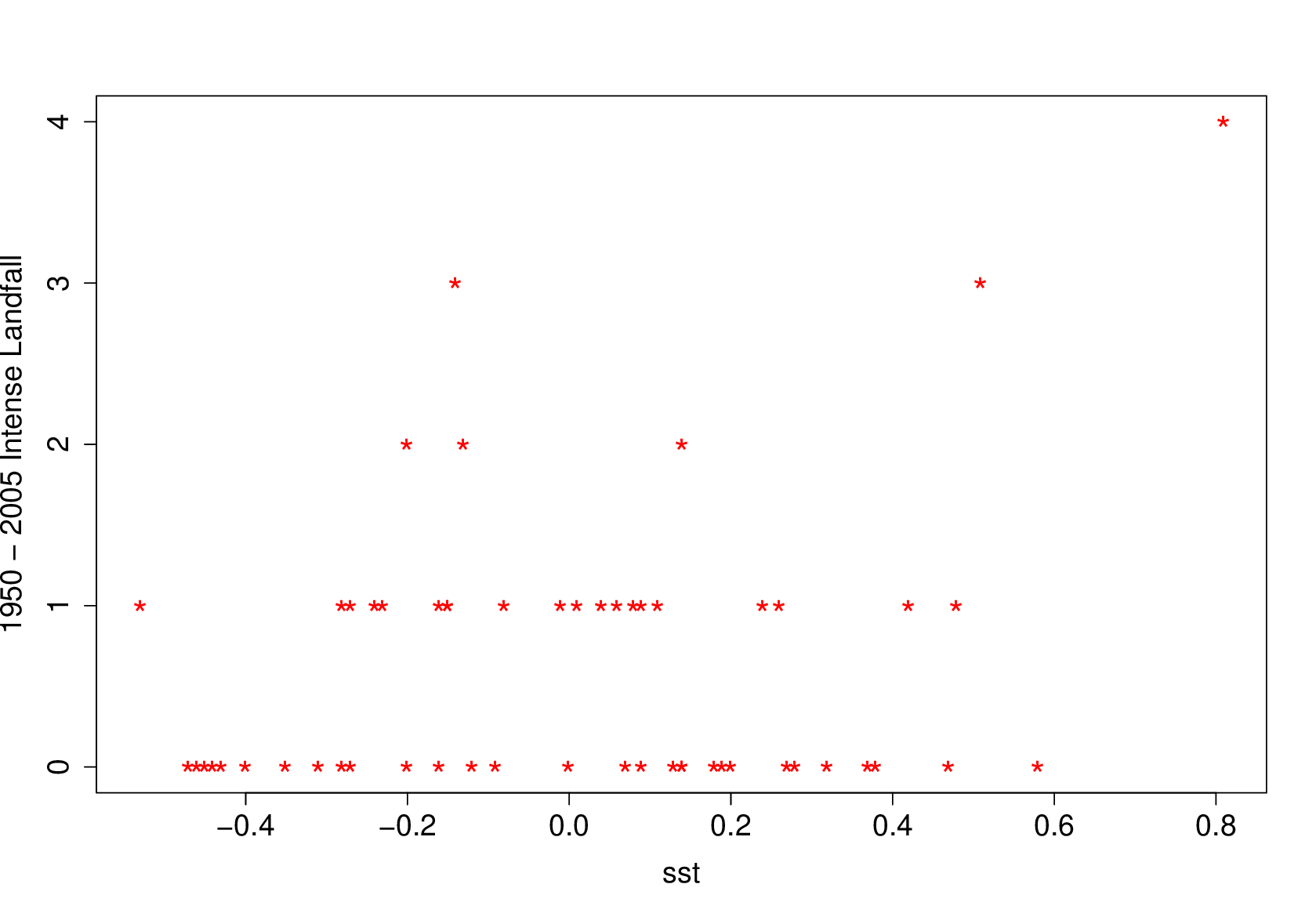}
} \caption{1950 - 2005 Intense Landfall vs. SST}
\end{figure}

\begin{figure}[h!]
\centering {
\includegraphics[width=8cm, angle=-90]{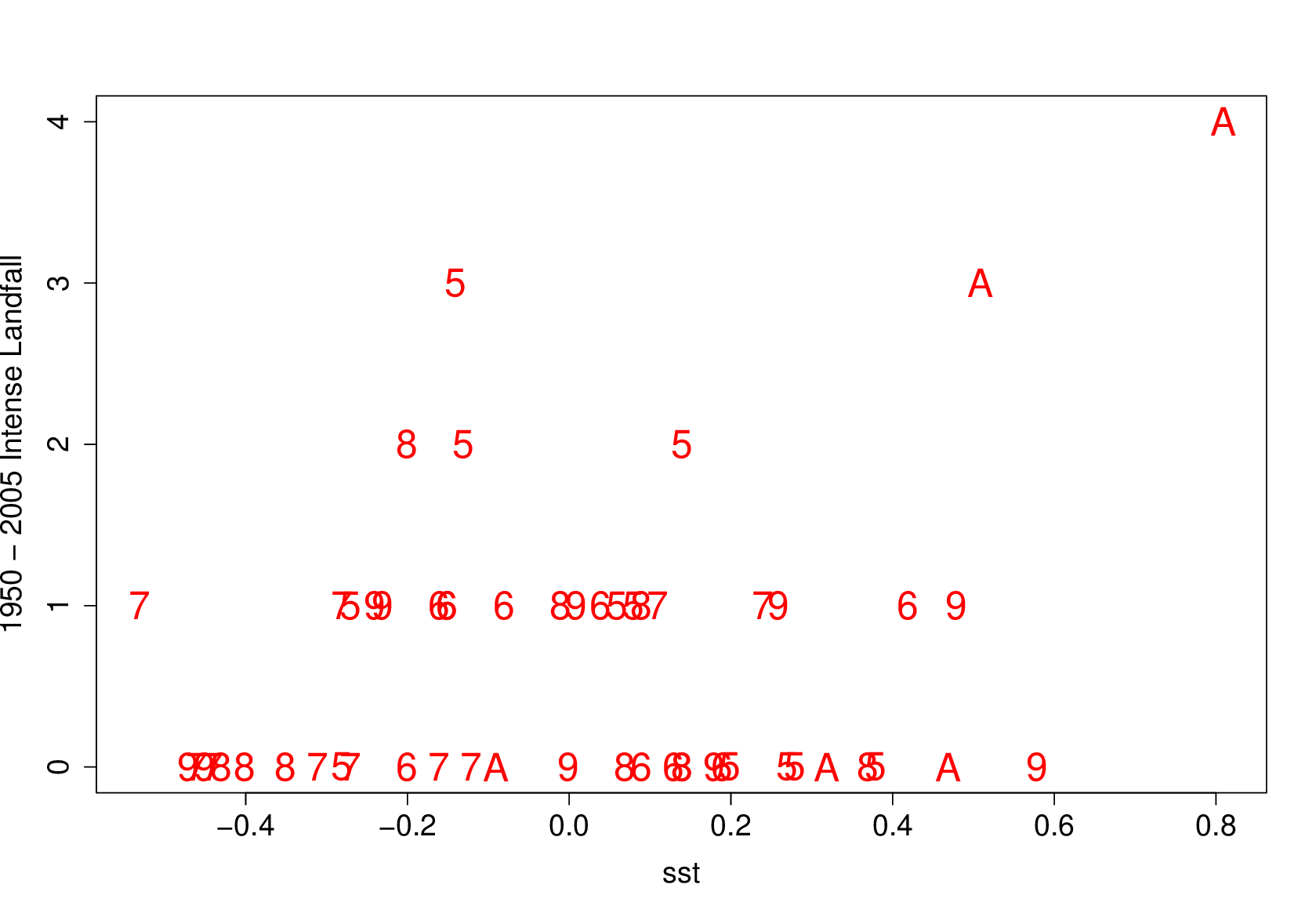}
} \caption{1950 - 2005 Intense Landfall vs. SST}
\end{figure}
\clearpage

\begin{figure}[h!]
\centering {
\includegraphics[width=10cm, angle=-90]{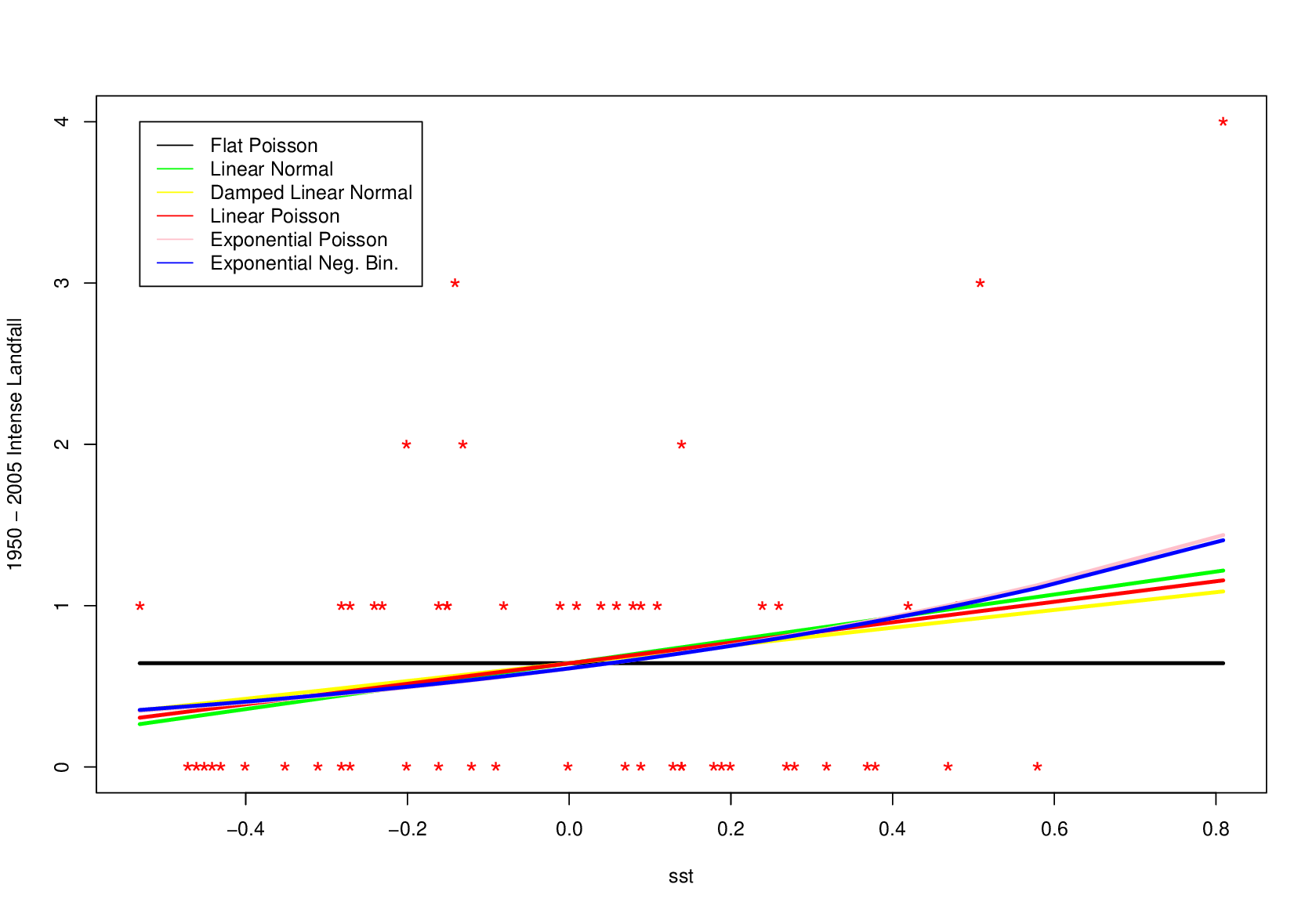}
} \caption{Fitted Lines for all Models 1950 - 2005 Intense
Landfall vs SST}
\end{figure}

\end{document}